\documentclass[12pt]{article}
\usepackage{graphicx}
\usepackage{epsfig}
\def\salto{\par\vskip .5cm}
\begin{document}
\begin{titlepage}

\title{INTRODUCTION TO BRS SYMMETRY}
%\REF{10}
\author{C. BECCHI\\  Dipartimento di Fisica, Universit\`a di Genova,\\
 Istituto Nazionale di Fisica Nucleare, Sezione di Genova,\\
  via Dodecaneso 33, 16146 Genova (Italy)\date{}}
\maketitle

\centerline{\bf Abstract}

This paper contains a revised version of the lecture notes of a short course on the quantization of gauge
theories. Starting from a sketchy review of scattering theory, the paper describes
the lines of the BRST-Faddeev-Popov quantization considering the problem of a
non-perturbative extension of this method. The connection between the Slavnov-Taylor
identity and the unitarity of the S-matrix  is also discussed.

 \vfill \footnote{Lectures given at
the ETH, Zurich, May  22-24, 1996 - December 2008 revised version }

\end{titlepage}

\def\.{\cdot}
\def\la{\lambda}
\def\s{\sigma}
\def\bs{{\bar\sigma}}
\def\t{\tau}
\def\o{\over}
\def\v{\vec}
\def\a{\alpha}
\def\z{\zeta}
\def\c{\gamma}
\def\b{\beta}
\def\d{\delta}
\def\k{\chi}
\def\f{\phi}
\def\C{\Gamma}
\def\S{{\bf S}}
\def\P{\Psi}
\def\T{\Theta}
\def\La{\Lambda}
\def\O{\Omega}
\def\x{\xi}
\def\n{\eta}
\def\u{\omega}
\def\ub{{\bar \u}}
\def\o{\over}
\def\p{\partial}
\def\ip{\int {d p\o  (2\pi)^4}}
\def\inx{\int d^4 x}
\def\iny{\int d^4 y}
\def\inz{\int d^4 z}
\def\dv{d \vec}
\def\ne{\not=}
\def\+{\bigoplus}
\def\D{\Delta}
\def\fo{{\cal F}_0}
\def\oo{{\cal O}}
\def\fc{{\cal F}_C}
\def\({\left(}
\def\){\right)}
\def\[{\left[}
\def\]{\right]}
\def\l.{\left.}
\def\r.{\right.}

\def\sec{\section}
\def\ss{\subsection}
\def\be{\begin{equation}}
\def\ee{\end{equation}}
\def\bea{\begin{eqnarray}&&}
\def\eea{\end{eqnarray}}
\def\nn{\nonumber \\ &&}
\def\nnn{\nonumber \\ }
\def\acca{\right.\nn\left.}
\def\ber{\begin{array}}
\def\eer{\end{array}}

\def\sp{\par\vfill\eject}
\def\quasisalto{\par\vskip .4cm}
\def\saltone{\par\vskip .7cm}
\def\saltino{\par\vskip .2cm}
\sec{Introduction}
\salto
These lectures begin recalling some general results of scattering theory 
\cite{1}.  The reduction formulae for the S-matrix  are given in terms of the Feynman
functional in the case of a massive field theory. Then the analysis comes to  gauge theories for which the concept
of gauge orbit is introduced.  The Faddeev-Popov definition \cite{fp} of a finite
functional measure is given. The BRST external differential operator along the orbits
is introduced together with the full  BRS operator \cite{brs}. The gauge algebra of the
infinitesimal gauge transformations is briefly discussed.
Assuming the existence of a global gauge fixing (no Gribov ambiguity \cite{3})
the Slavnov-Taylor identity and the gauge fixing independence of the
theory is deduced from the BRS invariance of the functional measure.
The extension of the Faddeev-Popov formula to the case of Gribov
ambiguities is briefly discussed together with that of the Slavnov-Taylor
identity.
The Slavnov-Taylor identity is then translated in terms of the proper
functional ( the effective action ) and the extension of the method to the
case of gauge algebras that are closed only modulo the field equations is
discussed.
Limiting for simplicity the study to the case of massive fields the
Slavnov-Taylor identity is applied to the two-point functions, this leads
to the introduction of the BRS symmetry for the asymptotic fields ( in
the form of Kugo and Ojima \cite{ko} )  and to the proof the existence of a
physical Hilbert space in which the S-matrix is unitary.

\sec{The S-matrix}
\salto
In quantum field theory \cite{1} the scattering amplitudes are 
computed by means of the reduction formula. This can be simply written using
the Feynman functional  generator of the theory that is defined according:
\be Z[j]\equiv e^{iZ_c}=<\O,T\( e^{i\inx \f (x) J(x)}\) \O>\ .\label{1}\ee
where $\f$ and $j$ in general label a set of quantized fields  and
corresponding sources. $Z_c$ is the
connected functional. The Feynman functional is computed by means of the 
formula: \be Z[j]=\int d\mu\  e^{i\inx \f (x)J(x)}\ ,\label{2}\ee in terms
of the functional measure $d\mu$ of the theory that is deduced from its bare
action by the heuristic relation; $d\mu=\prod_x d\f (x) e^{i S (\f )}$.

Computing the two-point function according:
\be{\d^2\o\d J_i(x)\d J_j(0)}Z_c|_{J=0}\equiv \D^{ij}(x)\ ,\label{3}\ee
and excluding for simplicity the presence of massless fields,  we can
separate from $\D$ the asymptotic propagator $\D_{as}$:
\be \D^{ij}(x)=\sum_\la\ip{e^{ipx}\o m_\la^2-p^2-i0_+}\zeta_\la^{ij}(p)
+R^{ij}(x)\equiv \D^{ij}_{as}(x)+R^{ij}(x)\ ,\label{4}\ee
where the Fourier transform of $R$ has no pole in $p^2$. It is clear that the
asymptotic propagator is by no means unique since $\zeta_\la^{ij}$ is defined up to a
polynomial in $p^2$ vanishing at $m_\la^2$; however this lack of uniqueness
does not affect the $\S$ matrix. Then one introduces the asymptotic free fields
$\f_{in}$ with the (anti-)commutation relations: \be\[\f_{in}^{i (+)} (x),\f_{in}^{j (-)}
(0)\]_{\pm}=\sum_\la\ip e^{ipx} \theta (p^0)\d (p^2-m_\la^2)\zeta_\la^{ij}(p)\ ,\label{5}\ee
and the asymptotic wave operator which satisfies: \be K_{ij}(\p)\f_{in}^{i} (x)=0\ 
,\label{6}\ee and \be  K_{ik}(-ip)\tilde\D^{kj}(p)|_{p^2=m_\la^2}=\d_i^j \label{6'}\ee for any $\la$. It should be clear that the asymptotic wave operator is by no means unique in much the same way as the matrix $\zeta$.

Then one computes the S matrix of the theory according:
\be \S =:e^{\inx\f^i (x) K_{ij}(\p ){\d\o\d J_j (x)}}:Z  |_{J=0}\equiv 
:e^\Sigma : Z
|_{J=0}\
.\label{7}\ee
\sec{Gauge invariance and BRS symmetry}
\salto
Now we come to the quantization of gauge theories. In this course we shall
disregard the crucial problem of the explicit non-perturbative 
construction of the theory, limiting our analysis to the formal and symmetry
aspects that should simplify the construction and characterize the solution of
the full quantum theory. We shall be often concerned with the functional
measure of the theory; avoiding any consideration of its actual definition we
shall indifferently pass from the Minkowskian form
 \be d\mu=\prod_x d\f (x) e^{i S (\f)}\label{mm}\ee to the Euclidean one
 \be d\mu=\prod_x d\f (x)e^{- S (\f )}\ .\label{me}\ee Furthermore, in order to
simplify the notation we shall merge all the labels of the fields into a single   
index including the space-time variable. Assuming the usual convention of
summation over repeated indices we shall also often omit the integration
symbol. However one should keep firmly in mind that the fields are local
variables and that locality is considered to play a crucial role in field
theory. For many reasons we shall also avoid discussing many mathematical
aspects that should bring our analysis too far from its purposes.
Let us call ${\cal F}_0$ the field space, that is the configuration space upon
which the gauge theory is constructed. In a gauge theory $\fo$ is fibered by the
gauge orbits ${\cal O}$ that is the set of gauge transforms of a given
configuration.  Considering the
infinitesimal transformations and translating everything in differential
geometry terms ( we are freely
following e.g. \cite{2}), we are given a system of partial differential
operators $\{X\}$ on $\fo$, that we shall label with the index $I$,
 and that in any point of
$\fo$ define a system of tangent vectors to the corresponding orbit. Denoting the
generic field coordinate in $\fo$ by the $\f^\a$,  we can write these operators 
in the form \be X_I=P^\a_I(\f )\p_{\f^\a}\ .\label{8}\ee

In the following we shall systematically use the symbol $\f^\a$, or, more explicitly $\f^\a(x)$, for the bosonic gauge and matter fields and we shall mention only occasionally the spinor fields. The bosonic gauge and matter field components are chosen real and hence they correspond to Hermitian operator-valued distributions.

The explicit form of Eq.(\ref{8}) in a pure non abelian case is:
\be X_I(x)=\p_\mu{\d\o\d A^I_\mu(x)}-gf_{IJ}^K  A^J_\mu(x) 
{\d\o\d A^K_\mu(x)}\ .           \ee 
The system $\{X\}$ is often called the differential
system of the orbits, whose very existence implies:
\be\[X_I,X_J\]=C_{IJ}^K (\f )X_K\ ,\label{9}\ee
that is the complete integrability of $\{X\}$. In the standard situation
the algebra (\ref{9}) is a Lie algebra, the structure functions $C_{IJ}^K$ are
constants. As a matter of fact, having merged the space-time variables with the discrete indices, what we call a Lie algebra is an infinite Lie algebra. Putting the space-time variables into evidence, that is, replacing above  $I, J$ and $K$ with the pairs $(I,x), (J,y) $ and $(K,z)$, $C_{IJ}^K$ decomposes into factors such as the constant $f_{IJ}^K$ and the distribution $\d(x-y)\d (x-z)$.
If the discrete indices run in the set $I=1,\dots, G$ we shall call $G$ the dimension of the gauge Lie algebra.

We shall see in the following how (\ref{9}) can be weakened
restricting the integrability condition to the "mass shell", that is modulo
the field equations. We also assume that a $X$-invariant measure is uniquely
defined up to an orbit-independent normalization constant. This we shall call the
vacuum invariant measure, the measure which is associated with the vacuum state of the theory. In 
general one is interested in the vacuum correlators of local observables; these
correspond to the integrals of a different class of invariant measures that can be
written as the product of the vacuum measure times  gauge invariant functionals
depending on the field variables corresponding to a suitably localized space-time
domain. A generic invariant measure will be assumed to belong to this class.

 The vacuum functional measure is constant over the orbits
${\cal O}$; in general this makes the functional measure of $\fo$ non integrable
and the Feynman functional ill defined. 
This difficulty is cured by the
Faddeev-Popov trick. In order to recall it conveniently let us assume that, even 
 if this is in general not the case, that the original
field variables trivialize the fibration; that is let us assume that the set
of fields $\{\f\}$ is decomposed according $\{\x\} $ and $\{\n\}$ where  $\{\x\}
$ are constant along the orbits and $\{\n\}$ are "vertical" coordinates along the orbits. Then
it is natural to make the measure integrable by multiplying it by an integrable
functional of $\n$ whose integral over ${\cal O}$ corresponding to the above
mentioned $X$-invariant  measure shoould be independent of the orbit (of $\xi$). One
often considers the invariant Dirac $\d_{inv} \[\n-\bar\n\]$:
\be \d_{inv} \[\n-\bar\n\]\equiv\d \[\n-\bar\n\] det |X_I\ \n^J|\ ,\label{10}\ee
 but more generally one can consider its
convolution  with a suitable $\bar\n$-functio-
nal.

Notice that the determinant appears in (\ref{10}) since the action of a gauge
transformation does not correspond to a Euclidean transformation on the $\n$
variables. The Faddeev-Popov measure \cite{fp} is obtained by the substitution:
 \be d\mu\rightarrow d\mu\ 
\d_{inv}\[\n-\bar\n\]\ , \label{11}\ee 
The invariant  Dirac measure can be
easily written as a functional Fourier transform.
Introducing two
sets of Grassmann variables $\{\u^I\}$ and $\{\ub_J\}$ that can be simply
identified with the generators of an exterior algebra  and the corresponding
derivatives that we label by $\{\p_{\u ^I}\}$ and by $\{\p_{\ub _J}\}$,
one introduces the Berezin integral: \be\int d\u^I\equiv{1\o\sqrt{2\pi i}}
\p_{\u ^I}\
,\label{11a}\ee for $\u$ and an analogous definition for $\ub$. Then using
the so called Nakanishi-Lautrup multipliers $\{b_J\}$ one can reproduce the
right-hand side of (\ref{10}) in the form: 
\be\int\prod db_I\prod d\u^J\prod d\ub_K\
\ e^{i\[ b_I\n^I- \ub_I\u^JX_J\n^I\]}\ .\label{13}\ee
This formula can be interpreted as an enlargement of the field space $\fo$
with the addition of a set of ordinary fields corresponding to the 
Nakanishi-Lautrup multipliers and of two sets of anticommuting fields
corresponding to the exterior algebra generators. We call $\fc$ the new space. We
also introduce the measure on $\fc$:
\be d\mu_C\equiv d\mu\prod_I db_I\prod_J d\u_J\prod_K d\ub_K \ .\label{mc}\ee

Looking now  into the details of  (\ref{13}), we see that  the
differential operator $\u^IX_I$ appearing in the exponent can be replaced
by its minimal nilpotent extension :
 \be
d_V=\u^IX_I-{1\o2}\u^I\u^J C_{IJ}^K (\f )\p_{\u_K}\ ,\label{15}\ee
 This
operator, that is often called the BRST operator \cite{brs}, is nilpotent due to
(\ref{9}) and to the corresponding Jacobi identity. That is:
 
 \bea\label{16}d_V^2=0 \leftrightarrow\left
\{\ber{l}\u^I\u^J\(X_IX_J-{1\o2} C_{IJ}^K (\f )X_K\)=0;\\ 
\u^I\u^J\u^K\( C_{IJ}^M (\f )C_{MK}^L (\f )-X_IC_{JK}^L (\f )\)=0. 
\eer\right. \label{16a}\eea

 Identifying the system $\{\u\}$ with that of the $X$-left-invariant forms
  we can interpret  the differential operator  $d_V$ as the
vertical exterior differential operator  on $\fo$, that is, with the operator on
$\fo$ that in any point is identified with the exterior differential operator on the
orbit (see Appendix A). 

Let us now come back to the trivializing coordinates, it is clear that these exist
globally only in very special cases, in particular when the corresponding fibration is
trivial. However in order to define a finite measure through Eq.(\ref{11}) it is sufficient to identify a
global section of $\fo$, if it exists,  intersecting every orbit in  a single point. This condition is equivalent
to the existence of a system of local
functionals $\{\Psi (\f)\}$ that we shall continue to label with the index $I$,
for which the Jacobian $det|X_I\Psi^J |$ does not vanish in the points where
$\Psi=\bar \Psi $ for some $\bar\Psi$. 

Assuming this condition, we shall replace in the exponent
in  (\ref{13}) the coordinate $\n^I$ by a generic functional $\Psi^I(\f)$ writing the
exponent as: \be i S_{GF}=i\[ b_I\(\Psi^I-\bar\Psi^I\) -\ub _I d_V \Psi^I\]\ ,\label{17}\ee
The above formula gives the definition of the gauge fixing action $S_{GF}$.

Eq.(\ref{17}) can be translated into a simpler form introducing a new exterior
derivative $s$ acting on the algebra generated by $\u$ and $\ub$ whose action on $\f$
and $\u$ coincides with that of $d_V$  and:
\be s\ub=b\ \ \ ,\ \ \ sb=0\ , \label{18}\ee
and hence
\be s=d_V+b_I\p_{\ub_I}\ .\label{19}\ee
It it clear that $s$ is nilpotent and that (\ref{17}) is written:
\be S_{GF}= s\[\ub_I\(\Psi^I-\bar\Psi^I\)\]\ ,\label{20}\ee
it is also obvious that $s$ commutes with the physical functional measure $d\mu$.
A further generalization of the measure, that includes also the convolutions 
of  (\ref{20}) with generic functionals of $\bar\Psi$, is obtained extending the
choice of $\Psi$ to $b$, $\u$ and $\ub$-dependent local functionals. In the
following we shall replace $\ub_I\(\Psi^I-\bar\Psi^I\)$ with a generic functional
$\Theta$ carrying the same quantum numbers. In the standard situation $\Theta$ is
a strictly local quadratic functional of $b$, that is,   the space-time integral
of a second order polynomial in $b$, independent of its derivatives; therefore $b$
is an auxiliary field. However there are models, in particular in supergravity, in
which \cite{nie} $b$ corresponds to propagating degrees of freedom that play the
role of extra ghosts. 

The most frequently met gauge choice in the case of renormalizable  theories is the linear choice in which the gauge fixing function in 
Eq. (\ref{17}) is a linear function. Showing explicitly the dependence on the space-time variables one sets:
\be \Psi^I=\int dx \[V^I_\a(\p)\f^\a+{\xi\o2}\d^{IJ}b_J\]\ ,\label{lingau}\ee where $\d$ is the Kronecker symbol and the matrix $V$ is a real linear function of the space-time  derivatives and has  maximal rank, that is, rank equal to the number of independent components of the $b$ fields.
Furthermore $\bar\Psi=0$. It follows that:
\be S_{GF}=\int dx \[b_I (V^I_\a(\p)\f^\a+{\xi\o2}\d^{IJ}b_J)-\bar\u_IV^I_\a P^\a_J(\f)\u^J\]\ .\label{linac}\ee

Even within the linear gauge  there are many possibilities among which one chooses depending on the calculation purposes. The most frequent choices are the covariant ones, typically Lorentz's gauge in which $V^I_\a(\p)\f^\a=k^I_J \p^\mu A_\mu^J$ and 't Hooft gauges, $V^I_\a(\p)\f^\a=k^I_J \p^\mu A_\mu^J+\rho^I_i\Phi_i$, where  $A_\mu^I$ is a gauge vector field and $\Phi_i$ is a scalar field. Among the non-covariant choice one often meets the ligh-like axial
gauge in which $V^I_\a\f^\a=k^I_J n^\mu A_\mu^J$ where $n$ is a light-like vector. However the covariant choices are the most convenient ones for a general discussion.

The above mentiones condition that $\det|X_I\Psi^J|$ does not vanish, that is, the condition for the gauge degrees of freedom to be completely fixed, implies that $\det |C_{0,J}^I|$, the determinant  of the Fourier transform of the field-independent part of $X_J V^I_\a \f^\a$,  does not vanish for a generic choice
of $p$. In the Lorentz gauge $C_{0,J}^I(p)=-k^I_J p^2$ and in 't Hooft's one $C_{0,J}^I(p)=-k^I_J p^2+\rho^I_it_J^{ij}v_j$ where $t_J^{ij}v_j$ is the in-homogeneous part of the scalar field gauge transformation. Discussing the $\bf S$-matrix unitarity we  shall assume a covariant choice with the further condition that $C_{0,J}^I(p)$ is real and the ghost part of the Lagrangian is formally Hermitian together with the ghost field $\u$, while the anti-ghost is anti-Hermitian. Since the equation $\det |C_{0,J}^I|=0$  is an algebraic equation of degree $G$ in $p^2$ whose solutions  in the semi-classical approximation are the ghost masses, we assume that all these solutions are real and positive and that the ghost particles are kinematically stable.

Thus under the standard assumption of a closed gauge algebra (\ref{9}) $S$ has the
following structure:
\be S=S_{inv}(\f ) +S_{GF}
+\int dx \[\c_\a\u^IP^\a_I(\f)+{1\o 2}\z_IC^I_{JK}\u^J\u^K\]\ .\label{so}\ee
In the case of renormalizable theories this structure is obliged by
the condition that the dimension of the action should be limited by that of space-time.

A further comment on the auxiliary role of $b$ and hence of $\ub$ is here
necessary. In this study we are tacitly considering the field space $\fc$ finite
dimensional, strictly speaking this is, of course, not true since every field
corresponds to an infinite number of variables; however one assumes that some
mechanism, e.g. some kind of regularization, renders finite the number of effective
degrees of freedom. With this proviso let us consider:
\bea \int\prod_I db_I\prod_Jd\ub_J\ e^{is\Theta}=lim_{\epsilon \rightarrow 0}
\int\prod_I db_I\prod_Jd\ub_J\ e^{is\Theta-\epsilon\sum_Kb_k^2}
\nn=i\ lim_{\epsilon \rightarrow 0}\int\prod_I db_I\prod_Jd\ub_J\int_0^1dt\
s\(\Theta\) e^{is\Theta t-\epsilon\sum_Kb_k^2}\nn=d_V\[i\ lim_{\epsilon \rightarrow
0}\int\prod_I db_I\prod_Jd\ub_J\int_0^1dt\ \Theta e^{is\Theta
t-\epsilon\sum_Kb_k^2}\] \ ,\label{exact}\eea where we have used the fact that the
Berezin integral of a constant gives zero. Eq.(\ref{exact}) shows that the Faddev-Popov
measure corresponds to the insertion of a $d_V$-exact factor into the functional
measure and the fields $b$ and $\ub$ are auxiliary in the sense that they allow the
explicit construction of this factor in local terms. Furthermore we see from 
(\ref{exact}) that the resulting measure on $\fo$ that is:
\be d\mu\int\prod_Kd\u^K\prod_I db_I\prod_Jd\ub_J\ e^{is\Theta}\ee
is an exact top form. It follows that its integral over a compact cycle, such as a
 gauge group orbit of a lattice gauge theory,
vanishes \cite{neu}\footnote{I thank M.Testa for calling my attention to this
reference}. This is due to the fact that on a cycle the gauge fixing equation $\Psi=\bar\Psi$ has
 an even number of solutions whose contributions to the above measure cancel
pairwise. However it should be clearly kept in mind that according to the Faddeev-Popov
prescription the functional integral should not cover the whole orbits but only a
compact subset of every orbit containing a single solution of the gauge fixing
equation. To be explicit let us consider the extreme example in which $\fo$ reduces to
a circle, a single $U(1)$ gauge orbit. Choosing $\Theta=\ub \ sin\ \varphi$, setting 
$s=i\u\p_{\varphi}+b\p_{\ub}$ and integrating over the whole space, one gets 
\be\oint d\varphi \int db d\u d\ub e^{i b sin\ \varphi +\ub\u cos\ \varphi} =-i\oint
d\varphi\d\(sin\  \varphi\)cos\ \varphi=0  \ ,\ee
while with the actual prescription, one has:
\be i\int_{-\epsilon}^{\epsilon} d\varphi \int db d\u d\ub e^{i b sin\ \varphi 
+\ub\u cos\ \varphi}=1\ .\ee

To conclude this section let us remember \cite{3} that in the case
of covariant and local gauge choices the condition
that $\Psi=\bar\Psi$ defines a global section of the orbit space does not hold true,
the situation is less clear for the so called axial gauges, that however suffer even
worst diseases \cite{gau}. We shall see how this difficulty can be overcome in
the situation in which $\fo$ can be divided into a system of cells $U_a$ in which one
can find for every cell a $\Psi_a$ defining a section in the cell.

\sec{The Slavnov-Taylor identity}
\salto

The particular structure of the functional measure allows an immediate proof, up to
renormalization effects, of the Slavnov-Taylor (S-T) identity. That is: for
any measurable functional $\Xi$ :
\be\int d\mu_C \ e^{iS_{GF}} s\ \Xi=0\ .\label{21}\ee
Indeed, using the same arguments as for (\ref{exact}), we get:
\bea\int d\mu_C \ e^{iS_{GF}} s\ \Xi=\int d\mu_C\  s\[ e^{iS_{GF}} \ \Xi\]\nn=
\int d\mu\prod_Id\u^Id_V\int\prod_Jdb_J\prod_Kd\ub_K\ e^{iS_{GF}} \ \Xi=0\
,\label{sti} \eea
since the last expression apparently corresponds to an exact top form whose support,
according to the general prescription, is contained in the integration domain, and
hence it vanishes on the boundaries of this domain.
 Considering   the extreme example given in the last section, one has for any continuous $A(\varphi)$:
\bea i\int_{-\epsilon}^{\epsilon} d\varphi \int db\ d\u d\ub e^{i b sin\ \varphi 
+\ub\u cos\ \varphi}s\[\ub A\(\varphi\)\]\nn=i
\int_{-\epsilon}^{\epsilon} {d\varphi} \int db\ d\u d\ub e^{i b sin\ \varphi 
+\ub\u cos\ \varphi}\[b A\(\varphi\)-i\ub\u A'\(\varphi\)\] \nn=\int_{-\epsilon}^{\epsilon} {d\varphi\o2\pi} \int db\   e^{i b sin\ \varphi 
+\ub\u cos\ \varphi}\[b A\(\varphi\)\cos\varphi-i A'\(\varphi\)\] \nn=
-i\int_{-\epsilon}^{\epsilon}  {d\varphi\o2\pi} {d\o d\varphi} \int db\  e^{i b sin\
\varphi } A\(\varphi\)=0
\ , \eea
since $ \int db\  e^{i b sin\ \varphi } A\(\varphi\)$ vanishes at
$\varphi=\pm\epsilon$.

 The identity (\ref{21}) can be interpreted saying that all correlators between 
elements of the image of $s$ and
$s$-invariants ones  vanish. Indeed, according to the definition given in section 3, the
$s$-invariant functionals can be considered
 to be generic local factors in the invariant measure
$d\mu$.
Considering the $s$ operator as the natural extension of $d_V$, the exterior
derivative corresponding to the gauge transformations, it is natural to
assume as a basic principle of gauge theories the identification of
observables with $s$-invariant functionals. Due to the nilpotency of $s$ this set
contains the image of $s$, whose elements, however, correspond to trivial observables
according to (\ref{21}). Therefore the non-trivial observables belong to the quotient
space of the kernel of $s$ versus its image, that its to the cohomology of $s$.

It remains to verify that the cohomology of $s$ is equivalent to that
of the vertical exterior differential operator $d_V$. Indeed consider the functional
differential operator:
\be \D\equiv-\(b_I\p_{b_I}+\ub_I\p_{\ub_I}\)\ .\ee
Let $P$ be the projector on the
kernel of $\D$, that is on the $\ub$ and $b$-independent functionals. It is apparent
that $\D$ and hence $P$, commute with $s$. Therefore a generic functional $X$ which is $s$-invariant, that is satisfying:
\be \(d_V +b_I\p_{\ub_I}\)X=0\ ,\label{kern}\ee
is the sum of two terms: $PX$ and $(1-P)X$, satisfying:
\be d_VPX=0\quad,\quad s\(1-P\)X=0\ .\ee
In much the same way, an  element of the image of $s$: $Z=sY$, is decomposed
according:
\be PZ=d_VPY\quad,\quad (1-P)Z=s(1-P)Y\ .\ee
Therefore the cohomology of $s$ is the union of that of $d_V$ in the kernel of $\D$
and that of $s$ in the kernel of $P$.
We want to verify that this second contribution is trivial. Indeed, consider the
differential operator: $\ub_I\p_{b_I}$, satisfying:
\be\{\ub_I\p_{b_I},s\}=-\D\ ,\ee
for $s$-invariant $X$, this yields:
\be s\ub_I\p_{b_I}(1-P)X=-\D(1-P)X\rightarrow 
(1-P)X=-s\D^{-1}\ub_I\p_{b_I}(1-P)X\ .\ee Indeed, on account of the definition of $P$, $\ub\p_\u(1-P)X$ belongs to the domain of $\D^{-1}$. Thus $(1-P)X$ belongs to the image of
$s$.

  A second consequence of (\ref{21}) is the gauge fixing independence of the
correlators of observables. Indeed let us compare
 the expectation values of the same
$s$-invariant functional $\O$ computed with two  different  measures corresponding to
choices of $\Theta$  differing by $\d\Theta$.
 To first order in $\d\Theta$ the difference of the expectation values is given by
\be\  i\int d\mu_C \ e^{iS_{GF}} s\( \d \Theta\) \O =0\ .\label{22}\ee
Of course this implies the independence of the expectation values of the choice of
$\Theta$ in a certain class of measurable functionals. Even in perturbation theory
this is not enough to prove that the expectation values in a renormalizable gauge
coincide with those in a non-renormalizable one.

We now come to the problem of extending the functional measure to the situation in
which the gauge fixing  is defined only locally.
In general the orbit manifold has to be divided into cells, each corresponding to a
different choice of $\T$.  Every cell in the orbit space corresponds to a cell $U_a$
in $\fo$. Let $\chi_a (\f)$ be a suitable smooth positive function with support in
$U_a$ and such that the set  $\{\chi\}$ is a partition of unity on the union of
the supports of the gauge-fixed measures $d\mu_C e^{is\T_a}$. This is shown in the
figure where the dotted lines corresponds to the support of the measures and the
circles to the cells.
\begin{figure}\begin{center}
\includegraphics*[scale={0.42}, clip=false]{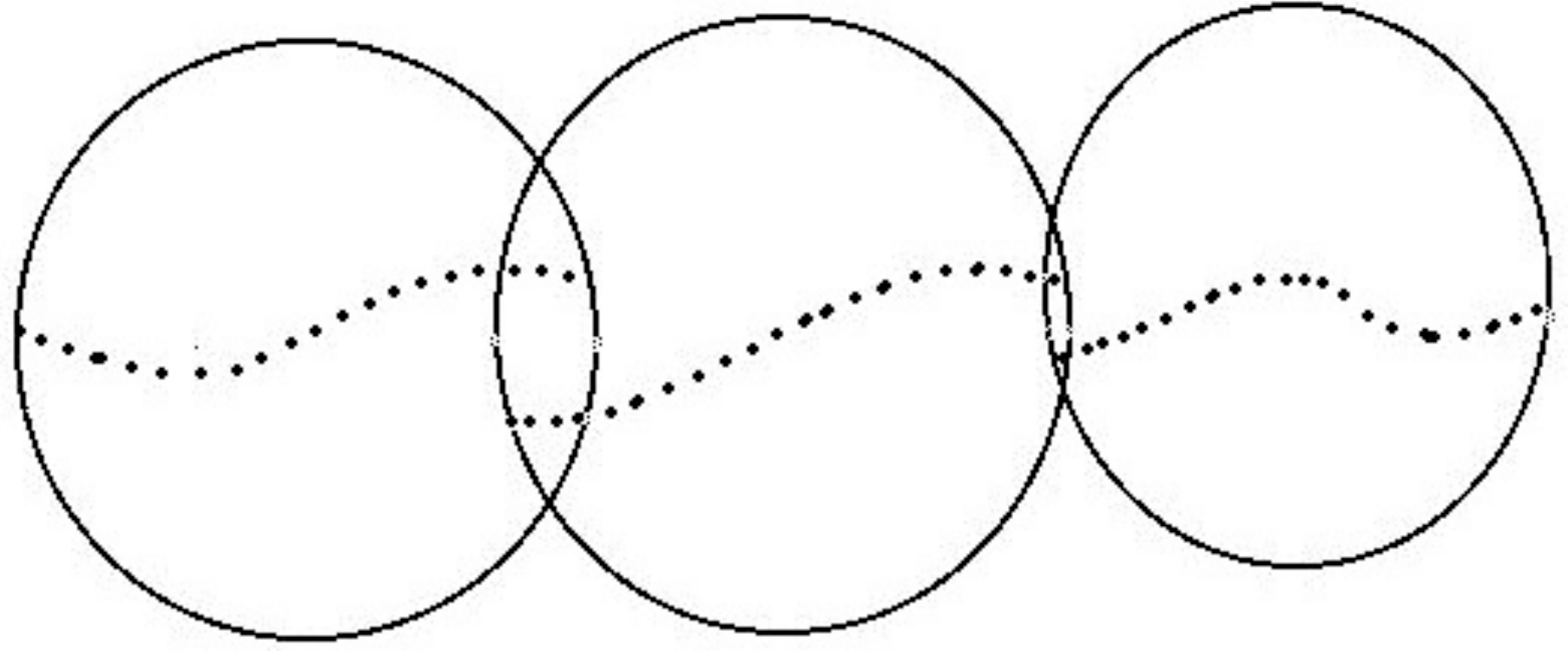}\end{center}
\end{figure}
  Explicitly a cell will be defined giving its center, that
is a special configuration $\f_a$ (background field), and defining the characteristic
functions $\chi_a$ according:\be
\chi_{a}(\f)\equiv{ \theta\( R^2-\Vert\f-\f_a\Vert^2\) \theta^2\( R^2-\mathrm{inf}_{c}\Vert\f-\f_c\Vert^2\)\over
\sum_{b}\theta\( R^2-\Vert\f-\f_b\Vert^2\)}.\label{pa}
\ee

%\be\chi_a(\f)=\theta\(R^2_a-\Vert\f-\f_a\Vert^2\)\prod_b\epsilon\(\Vert\f-\f_b\Vert^2-
%\Vert\f-\f_a\Vert^2\)\ ,\ee
where $\theta$ is a smoothed Heavyside function and $\Vert\f-\f_a\Vert$ is the $L^2$ norm of the difference
 $\f -\f_a$. Hints about the values of $R_a$ can be found in \cite{zw}. 

The BRS invariant functional measure corresponding to this
local gauge choice is given by \cite{bi}: \bea  d\mu_C\[\sum_a\chi_ae^{is\T_{a}}-
i^n\sum_{n=1}^{\infty}{(-1)^{n(n-1)\o2}\o n+1}\(s\chi_{a_{1}}...s\chi_{a_{n}}\
\chi_{a_{n+1}}\)_A
\r.\nn\l.\p_{\T}\(\T_{a_{1}}...\T_{a_{n+1}}\)e^{is\T_{a_{1},...,a_{n+1}}}\]\
,\label{1a}\eea 
where we have used the definitions:
\be \p_{\T}\(\T_{a_{1}}...\T_{a_{n}}\)\equiv\sum_{l=1}^n(-1)^{l+1}
\T_{a_{1}}...\check\T_{a_l}...\T_{a_{n}}\ .\label{def1}\ee
and:
\be
e^{is\T_{a_{1},...,a_{n}}}\equiv\int_0^\infty\prod_{i=1}^ndt_i\ \d\(\sum_{j=1}^nt_j-1\)
e^{is\sum_{k=1}^n t_k \T_{a_k}}\ .\label{def2}\ee
In  Appendix  B [(\ref{a7})] it is proven that under the hypothesis of a finite multiplicity of cell intersections the measure
(\ref{1a}) satisfies the Slavnov-Taylor identity.

 The lack of gauge invariance of the
characteristic functions of the cells induces new contributions to the measure
localized on the cell (regularized) boundaries. Of course the above measure could be
ill defined if the cells would accumulate around some singularity of $\fo$. This could
perhaps induce instabilities of BRS symmetry in the sense of \cite{fu}.

 Another version of the S-T identity
concerns the Feynman functional  involving the sources $j_\a$ of $\f^\a$,
$J^I$ of $b_I$, $\bs_I$ of $\u^I$ and $\s^I$ of $\ub_I$. The new functional is defined according:
\be Z\[\ j , J , \bs , \s\ \]\equiv\int d\mu_C \ e^{iS_{GF}} e^{i\int dx \[j_\a\f^\a+J^I
b_I+\bs_I \u^I+\s^I \ub_I\]}\ ,\label{23}\ee where we have explicitly shown the space-time integral symbol. This more explicit form will be essential for the discussion of unitarity.
 
The new form of the S-T
identity is a particular version of Eq>(\ref{21}), that is:
 \be \int d\mu_C \ e^{iS_{GF}} s\ e^{i\int dx  \[j_\a\f^\a+J^I
b_I+\bs_I \u^I+\s^I \ub_I\]}=0\ .\label{24}\ee It is  possible, exploiting
the nilpotency of $s$, to translate (\ref{24}) into a functional differential
equation for $Z$; this requires the introduction of further sources for the
composite operators generated by the action of  $s$ on the fields. On these are the source has to introduce the source
$\c_\a$ for $s\f^\a$ and $\z_I$ for $s\u^I$. These sources, which are often called anti-fields, appear in a further factor in the
functional measure: \be d\mu_C \ e^{iS_{GF}}\rightarrow d\mu_C \ e^{iS_{GF}}
 e^{i\int dx \[\c_\a s\f^\a + \z_I s \u^I\]}=d\mu_C \ e^{iS_{GF}}
 e^{-i\int dx\  s\[\c_\a \f^\a - \z_I  \u^I\]} \,\label{25}\ee which remains $s$-invariant due since $s$ is nilpotent.
Notice that the introduction of the sources for the fields and their variations has
enlarged the functional exterior algebra upon which the Feynman functional is defined. In
particular $\s , \bs$ and $\c$ are odd elements of this algebra. In the
following formulae many derivatives are in fact anticommuting derivatives, this
induces some obvious changes of sign.

Now, inserting the new measure into  (\ref{24}) we get:
\be \int d\mu_C \ e^{iS_{GF}}e^{i\int dx  \[\c_\a s\f^\a + \z_I s \u^I\]}
s\ e^{i\int dx  \[j_\a\f^\a+J^I b_I+\bs_I \u^I+\s^I \ub_I\]}=0\ ,\label{26}\ee 
that is:
\bea \int d\mu_C \ e^{iS_{GF}}e^{i\int dx  \[\c_\a s\f^\a + \z_I s \u^I\]}
\int dx \[j_\b s\f^b - \bs_J s \u^J \r.\nn\l. - \s^K b_K\]
 e^{i\int dx \[j_\a\f^\a+J^I b_I+\bs_I \u^I+\s^I \ub_I\]}=0\ .\label{27}\eea 
This is equivalent to the first order partial differential equation for the
extended Feynman functional:
\bea\int dx  \[j_\b  {\d\o\d\c_\b} - \bs_J {\d\o\d\z_J} - \s^K {\d\o\d J^K}\]
\int d\mu_C \ e^{iS_{GF}}e^{i\int dx  \[\c_\a s\f^\a + \z_I s \u^I\]}\nn
 e^{i\int dx \[j_\a\f^\a+J^I b_I+\bs_I \u^I+\s^I \ub_I\]}\nn\equiv {\cal S}Z=
\int dx \[j_\b  {\d\o\d\c_\b} - \bs_J  {\d\o\d\z_J} - \s^K {\d\o\d J^K}\] Z=0\ .\label{27a}\eea 
From now on, whenever we shall make explicit dependence on the space-time coordinates, we shall use for the functional derivative the  notation used in  Eq.(\ref{27a}).
This equation translates the S-T identity in terms of the Green functions.
The same equation holds true for the connected functional:
\be Z_c \[\ j , J , \bs , \s\ , \c , \z\]\equiv -i\  
log Z \[\ j , J , \bs , \s\ , \c , \z\]\ .\label{28}\ee
It is very useful to translate (\ref{28}) into a functional differential equation for the proper functional \cite{1}.
In perturbation theory the proper functional is the functional generator of the 1-particle-irreducible
amplitudes and   is generally defined as the Legendre transform of $Z_c$. It is
often called the effective action, although this name is also shared by completely
different objects. Introducing the collective symbol ${\cal J}$ for the field
sources ($  j , J , \bs , \s\ $), ${\cal K}$ for 
the other sources ($ \c , \z $) and $\Phi$
for the fields ($ \f , b , \u , \ub $), one defines the field functional:
\be\Phi\[{\cal J},{\cal K}\]\equiv {\d\o\d{\cal J}}Z_c\[{\cal J},{\cal K}\]-  {\d\o\d{\cal
J}}Z_c\[0,0\]\ ,\label{29}\ee then, assuming that the inverse functional ${\cal
J}\[\Phi,{\cal K}\]$ is uniquely defined, one has the proper functional:
\be\Gamma\[\Phi,{\cal K}\]\equiv Z_c\[{\cal J}\[\Phi,{\cal K}\],{\cal K}\]-
\int dx  {\cal J}\[\Phi,{\cal K}\]\(\Phi+ {\d\o\d{\cal J}}Z_c\[0,0\]\)\  .\label{31}\ee
It is easy to verify that:
\be {\d\o\d\Phi}\Gamma\[\Phi \[{\cal J},{\cal K}\],{\cal K}\]=\mp{\cal J}\
,\label{pm}\ee 
\be  {\d\o\d{\cal K}}\Gamma\[\Phi ,{\cal K}\]|_{\Phi=
\Phi \[{\cal J},{\cal K}\]}=  {\d\o\d{\cal K}}Z_c\[{\cal J},{\cal K}\]\ .\ee
Therefore:
\be  {\d\o\d\Phi(x)} {\d\o\d\Phi'(y)}\Gamma\[\Phi ,{\cal K}\]|_{\Phi=
\Phi \[{\cal J},{\cal K}\]}=\mp\[ {\d\o\d{\cal J}(x)} {\d\o\d{\cal J}'(y)}Z_c\[{\cal J},{\cal
K}\]\]^{-1}\ . \label{pm1}\ee
That is: the second field-derivative of $\Gamma$ gives the full wave operator.
Notice that the minus sign in (\ref{pm})  and (\ref{pm1}) refers to commuting fields while in the
anti-commuting  case one has the plus sign.

Using the above identities one can immediately write the S-T identity for the proper
functional:
\be\int dx \[ {\d\o\d\f^\a}\C {\d\o\d\c_\a}\C+ {\d\o\d\u^I}\C {\d\o\d\z_I}\C+b_I {\d\o\d\ub_I}\C\]=0\
.\label{stp}\ee
This identity is a crucial tool in many instances, we shall exploit it in the
analysis of unitarity, even more important is however its role in renormalization.

It is apparent from Eq.(\ref{linac}) that  the Lagrangian density depends on the field $b$  only through linear and  bilinear terms. Thus this field behaves like a free field and satisfies a linear equation of motion which is not affected by the perturbative corrections and in the functional language turns out to be:\be J^I+\[V^I_\a(\p) {\d\o\d j_\a}+\xi\d^{IJ} { \d\o\d J^J}\]Z_c=0\ ,\label{eb}\ee and hence:
\be { \d\o\d b^I}\C=V^I_\a(\p)\f^\a+\xi \d^{IJ}b_J\ .\label{eqb}\ee

Equation (\ref{eb}) combined with (\ref{27a}), which holds true also for $Z_c$, gives:
\be \s^I+V^I_\a (\p) {\d\o\d\c_\a}Z_c=0\ ,\ee that is:\be {\d\o\d\ub_I}\C+ V^I_\a(\p){\d\o\d\c_\a}\C=0\ .\label{equu}\ee
We shall use these equations in order to simplify the analysis of the $\bf S$-matrix unitarity.

A third interesting application of Eq.(\ref{stp}) is the search for generalizations of
the geometrical setting of gauge theories. This is based on the fact that  
(\ref{stp})
is verified by the classical action $S$ of a gauge theory. Indeed the classical
action is the first term in the loop-ordered perturbative expansion of $\C$.

The search for generalizations is justified by the fact that the low energy effective actions of more general theories, such as e.g.
supergravity, are free from dimensional constraints; this allows the introduction
of terms of higher degree in the sources $\c$ and $\z$. Disregarding the gauge 
fixing,  setting $\ub=b=0$, let us consider for example:
\be S=S_{inv}+\c_\a\u^IP^\a_I(\f)+{1\o 2}\u^I\u^J\c_\a\c_\b R^{\a\b}_{IJ}+
{1\o 2}\z_IC^I_{JK}\u^J\u^K\ ,\label{s1}\ee
that inserted into (\ref{stp}) gives:
\bea\u^IP_I^\a\p_\a S_{inv}=0\ ,\nn
 \c_\b\u^I\u^J\[P^\a_J\p_\a P^\b_J -
{1\o 2}P^\b_K C^K_{IJ}+ R^{\a\b}_{IJ}\p_\a S_{inv}\]=0\ ,\nn
\z_K\u^I\u^J\u^L\[P^\a_I\p_\a C^K_{JL}+ C^K_{MI} C^M_{JL}\]=0\ ,\nn
\c_\a \c_\b\u^I\u^J\u^L\[\p\c P^\b_I R^{\c\a}_{JL}+{1\o2}\(P_I^\c\p_\c
R^{\a\b}_{JL}+R^{\a\b}_{MI}C^M_{JL}\)\]=0\ ,\nn
\c_\a \c_\b\c_\c\u^I\u^J\u^K\u^L\[ R^{\d\a}_{IJ}\ \p_\d  R^{\b\c}_{KL}\]=0\nn
\c_\a \z_K\u^I\u^J\u^K\u^L\[ R^{\d\a}_{IJ}\ \p_\d  C^M_{KL}\]=0\ .\label{bv}\eea
In order to simplify the notation we have written $\p_\a$ instead of $\p_{\f^\a}$.
The first line above implies the invariance of $S_{inv}$ under the action of
the differential system ${X}$ given in (\ref{8}), the second one takes place of the first equation in
(\ref{16a}). It is indeed clear that the first term in this line corresponds to 
the commutator of two $X$'s, the second one prescribes the structure functions of the
algebra while the last one is new; it defines the deviation from a closed algebra
that, being proportional to the field derivative of the  physical action,
vanishes on the mass shell.
The third equation  in (\ref{bv})  prescribes the deformed structure of the Jacobi identity, that is, of the  the
second line of Eq.(\ref{16a}), while the remaining lines give constraints for $R^{\a\b}_{IJ}$. These constraints depend on the particular choice
of Eq.(\ref{so}) which excludes terms of second order in $\z$.
Eq.(\ref{so}) shows the simplest example of the extensions of our method to open
algebras that have been introduced by Batalin and Vilkovisky \cite{4}. 

A very simple example of a mass-shell closed gauge algebra can be found if one tries
to use the BRS algorithm to compute a $n$-dimensional Gaussian integral in polar
coordinates\footnote{This exercise has been suggested by J.Fr\"ohlich}. Let ${\vec
x}$ with components $x_i\ ,\ (i=1,..,n)$ be the variable and $S_{inv}=-{x^2\o2}$
define the invariant measure under the action of the gauge group $O(n)$ corresponding
to the BRS transformations:
\be sx_i=\u_{ij}x_j\, \label{ex1}\ee
where $\u_{ij}$ is the $O(n)$ ghost antisymmetric in its indices and, as usual,
 the sum over repeated indices is understood. The polar coordinate gauge choice
corresponds to the vanishing of $n-1$ components of  ${\vec x}$. This configuration
has a residual $O(n-1)$ invariance and therefore the Jacobian matrix in (\ref{10}) is
highly degenerate. To overcome this difficulty one has to enlarge the BRS structure
adding ghosts for ghosts ($\c_{ij}$); and hence introducing the ghost transformations:
\bea s\u_{ij}=\c_{ij}-\u_{ik}\ \u_{jk}\nn
s\c_{ij}=\c_{ik}\ \u_{jk}-\u_{ik}\ \c_{jk}\ .\label{ex2}\eea
With this choice $s$ is not nilpotent; indeed $s^2xi=\c_{ij}x_j$.
It is mass-shell nilpotent since the "field equations" are: $\p_{x_i}S_{inv}=x_i=0$.
Disregarding the structure of the gauge fixing and introducing a suitable set of
sources, we identify the form of the action (\ref{so}) which is  adapted to the present case according:
\be S=S_{inv}+\mu_is x_i +\z_{ij}s\u_{ij}+\eta_{ij}s\c_{ij}+{1\o2}\mu_i\mu_j\c_{ij}\
.\label{ex3}\ee
Eq.(\ref{ex3}) satisfies the Slavnov-Taylor identity (\ref{stp}) ( for $b=0$ ) The
 last term in Eq.(\ref{ex3}) corresponds to the second order term in $\c$ in
Eq.(\ref{so}).

 \sec{Unitarity} \salto
 The first step in the analysis of $\bf S$ unitarity is the study of the
asymptotic propagators and wave operators of a gauge theory \cite{be}.  

For simplicity we shall limit our study to the situations in which the whole gauge symmetry is spontaneously broken and hence no  vector particle is left massless. We shall also assume that, contrary to the Electro-Weak model, all physical and unphysical particles are stable, and we choose a gauge fixing prescription of the 't Hooft kind in which no unphysical particle, in particular Faddeev-Popov ghost,  is mass-less. 

First of all let us now consider  how one can extract information about asymptotic particle states from the second derivatives of the proper functional. We label by $\Phi_i$ the fields appearing in our theory, that is the variables upon which $\C$ depends. In our case the index $i=1,\cdots, N$ distinguishes vector field components from ghost and matter field ones. We choose  the fields $\Phi$ Hermitian, with the exception of the anti-ghost which are chosen anti-Hermitian.
Defining the Fourier transformed field:
\be\tilde\Phi_i(p)\equiv\int {d^4x\o(2\pi)^4}e^{-ip\cdot x}\Phi_i(x)\ ,\label{four}\ee and setting:
\be\p_{\tilde\Phi_i(p)}\equiv \int d^4x\ e^{ip\cdot x}{\d\o\d\Phi_i(x)}\quad\ ,\quad
\p_{\Phi_i}\equiv {\d\o\d\Phi_i(0)}\ ,\label{def}\ee
we get the Fourier transformed wave matrix of the theory whose elements are the second derivatives:
\be \p_{\tilde\Phi_i(p)}\p_{\Phi_j}\C|_{\Phi=0}\equiv \C_{i,j}(p)\ .\ee With the exception of the ghost field components this is in general an Hermitian matrix. Due to translation invariance one has $ \C_{i,j}(p)=\pm \C_{j,i}(-p)$ where the upper  and lower signs refer  to commuting  and anti-commuting fields respectively.

For simplicity we limit our discussion to the cases in which  the determinant $\Delta(p^2)\equiv\det|\C(p)|$ is an analytic function of $p^2$ for $p^2<M^2$ and its  zeros in the analyticity domain lie on the positive real axis. 

Let  $(\bar p_\la)^2=m_\la^2$ for $\la=1,\cdots,a$  be   solutions of order $k_\la<N$ of the equation $\D( p^2)=0$. We assume that the matrix equation $\C(-\bar p_\la)_{i,j}v_j(\bar p_\la)=0$ has  $k_\la $ independent, non-trivial  solutions. That is, the matrix $\C(-\bar p_\la)$ has $k_\la$ independent null eigenvectors.   

This is a simplifying hypothesis which excludes the presence of dipole, or even worst, singularities in the unphysical propagators. These singularities are often met in gauge theories with particular gauge choices, e.g. QED in the Landau gauge and in the example we shall present in the following.
As a matter of fact it is shown in Appendix C that dipole singularities correspond to mass-degenerate asymptotic states with opposite norm, that is, they correspond to a pair of simple poles with opposite residue and degenerate in mass. Thus our analysis is easily extended to the dipole case, however here we prefer to begin considering the simplest case of simple poles.

Under the above assumptions near a zero $\bar p_\la$ of the determinant one has:
\be\C(p)=(p^2-m_\la^2)  \tilde\zeta_\la(p)+R_\la(p)\ ,\label{spectral}\ee
where the matrices $ \tilde\zeta_\la$ and $R_\la$ are Hermitian and by no means unique. As already noticed in section 2 this lack of uniqueness does not affect the scattering theory. Still, in a suitable neighborhood of $\bar p_\la$, we make the choice $ \tilde\zeta_\la R_\la=R_\la \tilde\zeta_\la=0$.

With this choice $ \tilde\zeta_\la(p)$ has rank  $k_{\la}$   and, if $P_\la(\bar p_l)$ is the projector upon the space spanned by the null eigenvectors of  $\C(-\bar p_\la)$, one has:  \be P_\la(p) \tilde\zeta_\la(p)= \tilde\zeta_\la(p)P_\la(p)= \tilde\zeta_\la(p)\ee Furthermore another matrix $\zeta_\la(p)$ with  rank $k_\la$ exists  such that $ \tilde\zeta_\la(p)\zeta_\la(p)=\zeta_\la (p) \tilde\zeta_\la(p)=P_\la(p)\ .$
Therefore one has:
\be \C(p)P_\la(p)=(p^2-m_\la^2) \tilde\zeta_\la(p)+ O((p^2-m_\la^2)^2)\ ,\ee and hence:\be  P_\la(p)=(p^2-m_\la^2)\C^{-1}(p) \tilde\zeta_\la(p)+ O((p^2-m_\la^2))\ , \ee and \be \C^{-1}(p)={\zeta(p)\o p^2-m_\la^2 -i0_+} +Q_\la(p)\ ,\label{propa}\ee where $Q_\la$ is analytic in the mentioned  neighborhood of $\bar p_\la$. This last equation leads to Eq.(\ref{4}) and hence allows the construction of the scattering matrix with the asymptotic fields satisfying the (anti-)commutation relations (\ref{5}).

In particular the expression for the scattering matrix, Eq.(\ref{7}), can be written in the alternative form involving the Fourier transformed asymptotic fields $\tilde\f_{in}$ and corresponding sources $\tilde J$:
\be {\bf S}=:\exp\int {d^4p\o (2\pi)^4}\tilde\Phi_{in}^i(p)\C_{j,i}(-p){\d\o\d\tilde J_j(p)}:Z|_{j=0}\ ,\label{scatte}\ee where,  taking into account Eq. (\ref{5}), we have replaced the asymptotic wave operator $K$  with the proper two point function $\C$.

 Now we consider how the Slavnov-Taylor identity for the proper functional  (\ref{stp})  and Eq.s (\ref{eqb}) and Eq.s (\ref{equu}) constrain the wave matrix of our gauge theory.
Setting: 
\be  {\d^2\o\d{\tilde \f}^\a (p) \d \f^\b(0)} \C |_{\Phi=0}\equiv \C_{\a\b}(p)\
,\label{d1}\ee  and
\be  {\d^2\o\d\u^I(0)\d{\tilde \c}_\b (p)}  \C |_{\Phi=0}\equiv \C_I^\b (p)\
,\label{d2}\ee  
and computing  in  the origin of the field manifold the second functional derivative  of Eq.(\ref{stp}) with respect
to ${\tilde \f}^\a (p)$ and $\u^I(0)$ we get:
\be \C_{\a\b}(p)\ \C^\b_I(p)=0\ .\label{trans}\ee  In QED this corresponds to the
transversality condition for the vacuum polarization. 

In general from Eq's (\ref{8})  and (\ref{so}) we see, up to quantum corrections,  that Eq. (\ref{d2})  defines $G$ vectors $\C_I(p)$,  where  $G$ is the dimension of the gauge Lie algebra mentioned in section 3. These vectors span the tangent space to the orbits in the origin. Therefore they must be independent   and they must remain such beyond the quantum corrections. 
 Eq. (\ref{trans}) shows that, for a generic $p$ the matrix $\C$ has $G$ null eigenvectors and hence its determinant  is identically zero.  We shall call $\C$   a {\it degenerate matrix}. For a generic choice of the momentum $\C$'s rank is equal to  $F-G$ where $F$ is the number  of its rows and columns .

From Eq.(\ref{eqb}) one get:
\be  {\d^2\o\d{\tilde \f}^\a (p)\d b_I(0)} \C|_{\Phi=0}=  V^I_\a (ip)  \ ,\label{d3}\ee
  thus setting, as above:
\be \  {\d\o\d\u^I(0)\d{\tilde \ub}_J (p)}\C|_{\Phi=0}\equiv \- C_I^J (p)\
.\label{d4}\ee and 
taking the second derivative of (\ref{stp}) with respect to ${\tilde b}_I (p)$ and $\u^J$, we get
\be \C^\a_J (p) V^I_\a (-ip)= -C^I_J(p)\ .\label{degene}\ee

We have furthermore from (\ref{eqb}):
\be  {\d^2\o\d{\tilde b}_J (p)\d b_I(0)} \C|_{\Phi=0}\equiv  \xi\d^{IJ}   \ .\label{d5}\ee The wave matrix of gauge and matter fields is then given by:
 \be\label{wavef}
\left(\ \ \ber{cc}\ \ \C_{\a\b}&\  V_\a^{I } \\ V _{\b}^{J*} &\ \  \xi\d^{I,J}
 \eer\ \right)(p)\ ,\ee where, taking account that $V(\p)$ is a real linear matrix, we have set $V_\a^{I } \equiv V_\a^{I }(ip)$ and $V_\a^{I *}=
 V_\a^{I }(-ip)$.
 
The wave operator of the Faddeev-Popov ghosts has matrix elements $C_I^J(p)$ . 

Let us now look for a null eigenvector of the above gauge and matter field wave matrix. We have to solve the system $\C_{i,j}(-p)\Phi^j(p)=0$, that is:
\bea \C_{\a\b}(-p)\f^\b(p)+V_\a^{I } (-ip)b_I(p)=0\nn V^{I}_\b(ip) \f^\b(p)+\xi b^I(p)=0\ \label{siste}\eea which implies:
\be \[\C_{\a\b}(-p)-{1\o\xi}V_\a^{I}(-ip) V_\b^{I }(ip) \]\f^\b(p)\equiv \bar\C_{\a\b}(-p)\f^\b(p)=0\  .\label{siste2}\ee and \be \xi b_{in}^I(p)=-V^I_\a(ip)\f_{in}^\a(p)\ .\label{siste3}
\ee
If the gauge fixing procedure is complete, the matrix $\bar\C$ must be non-degene-
rate, and the solutions to the equation $\det|\bar\C|(p^2)=0$ correspond to the masses of the bosonic gauge and matter fields. Notice that the spinor degrees of freedom are not considered since they only carry physical degrees of freedom.

Now, taking into account Eq.(\ref{trans}), Eq.(\ref{degene}) and Eq.(\ref{siste2}) we get:\be\bar\C_{\a\b}(p)\C^\b_I(p)={V_\a^{J}(ip)C_I^J(p)\o\xi}\ .\label{sti2} \ee
This equation relates the masses in the ghost sector to those in the gauge matter sector.
 The matrix $\bar\C$ is Hermitian  and hence it has the structure appearing in Eq.(\ref{spectral}).

If $\bar p_g$ is a solution  of the equation $\det|C|=0$, the linear system \be C^I_J (-\bar p_g)w^J(\bar p_g)=0\label{g1}\ee has non trivial solutions. Let $M_g$ be the number of the independent solutions of (\ref{g1}) that we label by $w_a^J(\bar p_g)$ for $a=1,\cdots, M_g$. From (\ref{sti2}) we have \be \bar\C_{\a\b}(-\bar p_g)\C^\b_I(-\bar p_g)w_a^I(\bar p_g)\equiv \bar\C_{\a\b}(-\bar p_g)\f^\b_a(\bar p_g)=0\ .\label{g2}\ee The $M_g$ vectors $\f_a$ are linearly independent since the vectors $\C_I$ are linearly independent. In other words, if one had a  non-trivial set of coefficients $c^a$ such that $c^a\f_a^\a =c^aw_a^I(\bar p_g)\C^\a_I(-\bar p_g)=0$, this would contradict, either the hypothesis of linear independence of the $w_a$'s, or that of independence of the $\C_I$'s.

In the same situation also the  linear system \be \bar w_I(\bar p_g) C^I_J (\bar p_g)=0\label{g3}\ee has $M_g$ linearly independent solutions that we label by $\bar w^a(\bar p_g)$. For every solution of $\det|C|(p)=0$ one has the same number of solutions of  Eq.(\ref{g1})  and of Eq.(\ref{g3}). 

Notice that, if there are solutions to Eq.(\ref{siste2}) with $b^I\not\equiv 0$, they correspond to solutions of Eq.(\ref{g3}). Indeed under the above condition $V_\a^I(ip)\f^\a(p)\not\equiv 0$, however one has:\be\C_I^\a(p)\bar\C_{\a\b}(-p)\f^\b(p)=-C_I^J(p)V_{\b}^J(ip)\f^\b(p)=\xi
C_I^J(p)b_J(p)=0\ .\ee These solutions with non-vanishing $b$ do not necessarily exist, indeed, as it will be shown in a moment, they may be replaced with dipole singularities. In any case the number of  solutions of Eq.(\ref{siste2}) with independent non-vanishing $b$ cannot exceed $G$. Let it be $G-X$. We claim that Eq.(\ref{siste2}) has $X$ independent dipole solutions.

In order to prove this claim we better specify our framework. We are considering models in which the bosonic fields are either gauge vector or scalar fields. Among the scalar fields there are $G$ Goldstone bosons associated with the spontaneous breakdown of gauge symmetry and required by the Higgs mechanism, and the same number of vector fields. Indeed in this situation and with the 't Hooft gauge choice, one can exclude massive asymptotic fields. Furthermore the bosonic wave operator is the sum the space-like vector field operator, that of the Higgs scalar fields, and the {\it gauge fixing} wave operator for the mixed the Goldstone fields and scalar components of the vector fields, that is $\p^\mu A_\mu$. We concentrate on this wave operator for the moment forgetting the rest.
With our Hermitian choice of the field basis the Fourier transform of the gauge fixing term of the wave operator defined in Eq.(\ref{d1}) is a real $p$-dependent matrix which has the following $2\times 2$ block structure:\
 \be\label{waveg}
\left(\ \ \ber{cc}\ \ \C_{11}&\  \C_{12} \\ \ \C^T_{12} &\ \  \C_{22}
 \eer\ \right)\ ,\ee where the suffix $T$ means "transposed". The blocks are $G\times G$ matrices and in particular $\C_{11}$ is the restriction of the wave matrix to the $\p^\mu A_\mu\ $ components and $\C_{22}$ that to the Goldstone bosons. According to Golstone's theorem 
$ \C_{22}=p^2K$ and, in our framework, $K$, which is a real symmetric matrix, is also invertible, thus we can write:\be   \C_{22}= O^Tp^2  d\ O\ee where $d$ is a real, diagonal and strictly positive matrix  and $O$ an orthogonal one.

The matrix $\C^\b_I(p)$ defined in Eq.(\ref{d2}) has the corresponding block structure:
\be\left(\ \ber{c} \c_{1} \\ \c_2 
 \eer\ \right)\ee where, due to the assumption of complete symmetry breakdown,  both $\c_1$ and $\c_2$ must be invertible matrices.
 
 Eq. (\ref{trans}) is written in terms of the above block matrices in the form:
 \be \C_{11}\c_1
+\C_{12}\c_2=0\quad\ ,\quad\C^T_{12}\c_1
+\C_{22}\c_2=0\ .\label{siste4}\ee
Changing the basis of the Goldstone bosons one can transform $\C_{22} $ into the unity matrix leaving 
$\C_{11}$ and $\c_1$ unchanged and transforming $\C_{12}$ into $\C_{12}O^T\sqrt{1/(p^2 d)}\equiv \C'_{12}$ and $\c_2$ into $\sqrt{p^2d}O \c_2\equiv \c'_2$. After this transformation Eq.(\ref{siste4}) becomes:
\be \C_{11}\c_1
+\C'_{12}\c'_2=0\quad\ ,\quad\C^{'T}_{12}\c_1
+ \c'_2=0\ ,\label{siste5}\ee which is solved by:\be \C'_{12}=-(\c_1^{-1})^T\c^{'T}_2\equiv R\quad\ ,\quad
 \C_{11}=R R^T\ .\ee The gauge fixing wave matrix $\bar\C$ defined in Eq.(\ref{siste2}) can be written as the difference $\bar\C=\C-\Delta$ where $\Delta$ is deduced from Eq.(\ref{siste2}). Here we generalize the gauge choice replacing the matrix $\xi\d_{IJ}$ with the symmetric real and invertible matrix $\xi^{IJ}$ getting the new matrix $\Delta$ in the block form:
 \be \left(\ \ \ber{cc}\ \ \xi^{-1}&\  \xi^{-1}V^T \\ \ V \xi^{-1}&\ \ V \xi^{-1}V^T
 \eer\ \right)\ .\ee With the new choice of the Goldstone field basis $\Delta$ keeps the same form with $V$ replaced with $\sqrt{1/(p^2 d)}\ O\ V\equiv W$. 
 
 The ghost wave matrix is computed from Eq.(\ref{degene}) getting:\be C=-\c^T_1-\c^T_2V =-\c^T_1(I-RW)\ .\label{gw}\ee
 Therefore, in order to study the masses of the  asymptotic fields corresponding to the Goldstone bosons and scalar components of the gauge fields, we have to compute the determinant of the matrix whose block structure is:
   \be \bar\C\equiv\left(\ \ \ber{cc}\ RR^T- \xi^{-1}&\  R- \xi^{-1}W^T \\ \ R^T-W \xi^{-1}&\ \ I-W \xi^{-1}W^T
 \eer\ \right)\ .\label{matri}\ee We shall use the following formula for a matrix in block form:
  \be\det \left(\ \ber{cc}\ A&\  B \\ C&\ \ D
 \eer\ \right)=\det(A-BD^{-1}C)\det D\ ,\label{formu}\ee where, of course, we have assumed that $\det D$ does not vanish\footnote{I thank C. Imbimbo for suggesting this
particular equation among many equivalent forms}. In order to exploit this formula let us consider the matrix:
\bea RR^T- \xi^{-1}-( R- \xi^{-1}W^T)( I-W \xi^{-1}W^T)^{-1}(R^T-W \xi^{-1})\nn=RR^T- \xi^{-1}-( R- \xi^{-1}W^T)\sum_{n=0}^\infty (W \xi^{-1}W^T)^{n}(R^T-W \xi^{-1})\nn=-(I-RW)\xi^{-1}\sum_{n=0}^\infty (W^TW \xi^{-1})^{n}(I-W^TR^T)\nn=--(I-RW)\xi^{-1}W^T(W^{-1})^T( I-W \xi^{-1}W^T)^{-1}(I-W^TR^T),\label{detto}\eea
where we have assumed the convergence of the operator power series. This does not limit the generality of our analysis since the operator series certainly convergences for a suitable choice of the matrix $\xi$ and, from Eq.(\ref{formu}) we get:
\be\det\bar\C=(\det\xi)^{-1}(\det C)^2/(\det\c_1)^2\label{dipo}\ee which is independent of $\det( I-W \xi^{-1}W^T)\equiv \det D$.

In order to simplify the discussion let us now assume that on a given ghost mass-shell, say $p^2=\bar m_0^2$, there is a single ghost solution $w_0$ (Eq.(\ref{g1})) and a single anti-ghost one $\bar w_0$ (Eq.(\ref{g3})). This hypothesis  does not limit the generality of our discussion since it can always be met through a suitable choice of the ghost fixing parameters. Our formula (\ref{dipo})   shows that on  the same mass-shell $\det\bar\C$ has a double zero Therefore, either Eq.(\ref{siste2}) has two solutions, that is,
$\hat\f^\a=\C^\a_I(-\bar p_0)w^I_0$ and a solution with $b_I=\bar w_{0,I}$, or the first solution is unique and the corresponding asymptotic field propagator has a dipole singularity. It is shown in Appendix C that in  this case the positive frequency part of the asymptotic field creates two degenerate particle states with opposite norm.
 
 The analysis of $\S$-matrix unitarity can be pusher forward in both situations, it is however simpler in the case of two solutions  \cite{brs}. Therefore, in the following, we shall consider this particular case.
%

%Let us define $\C_0^\a (p)\equiv \C^\a_I(p)w^I_0(-\bar p_0)$ and $V^0_\a(p)\equiv V^I_\a(ip)\bar w_{0,I}+...$ on the ghost mass-shell these two vectors are mutually orthogonal and they can be organized in such a way as to reduce the matrix $\bar \C$. The reduced matrix is two dimensional and hence, in the neighborhood  of the ghost mass-shell its determinant is a polynomial of second degree in $x\equiv p^2-m_0^2$.

Let us now compute, in the framework of our gauge theory, the operator $\Sigma$ appearing and defined in Eq.(\ref{7}). Limiting the analysis to the gauge and bosonic matter fields and to the ghost fields, that is, disregarding the spinor fields, one has:
\bea \Sigma_B=\int {d^4p\o (2\pi)^4}\[\f_{in}^\a(p)[\C_{\b\a}(-p){\d\o\d {\tilde j}_\b(p) }+V^{I }_\a(ip){\d\o\d {\tilde  J}^I(p) }]\acca+b_{in}^I(p)[V^{I}_\b(-ip){\d\o\d {\tilde  j}_\b(p)}+\xi{\d\o\d {\tilde  J}^I(p)}]-\u_{in}^I (p)C_I^J(-p){\d\o\d  \tilde{ \bar \s}^J(p)}\acca+\bar\u_{in,J}(p)C_I^J(p){\d\o\d  {\tilde \s}_I(p)}\]\ .\eea
Using Eq.(\ref{siste3}), we can transform this equation into the simpler form:
\bea  \Sigma_B=\int {d^4p\o (2\pi)^4}\[\f_{in}^\a(p)\bar\C_{\b\a}(-p){\d\o\d  {\tilde j}_\b(p)} -\u_{in}^I(p)C_I^J(-p){\d\o\d {\tilde  \s}^J(p)}\acca+\bar\u_{in,J}(p)C_I^J(p){\d\o\d \tilde  {\bar\s}_I(p)}\]\ .\label{sigb}\eea 
It is clear that the operator $\Sigma_B$ acts on the asymptotic states containing ghost and bosonic particles, since we are forgetting spinors. Among the bosonic particles there are those corresponding to the asymptotic fields $\f_a^\a(p)=\C_I^\a(-p)w_a^I(p)$ considered above in Eq.(\ref{g2})  and the asymptotic components of $b$. There also are the particles corresponding to the remaining independent solutions of Eq.(\ref{siste2}) that we call {\it physical} together with the spinor particles.

Computing the commutator $[{\cal S}, \Sigma_B]Z$ where;
\be {\cal S}\equiv\int dx \[j_\b {\d\o\d \c_\b }  - \bs_J {\d\o\d\z_J} - \s^K {\d\o\d J^K}\] \ ,\label{sop}\ee is the functional differential operator appearing in Eq.(\ref{27a})  and $Z$ is the Feynman functional, we get:
\be [{\cal S}, \Sigma_B]Z=-\int {d^4p\o (2\pi)^4}\[\f_{in}^\a(p)\bar\C_{\b\a}(-p){\d\o\d \tilde  \c_\b(p)} +\u_{in}^I(p)C_I^J(-p){\d\o\d {\tilde  J}^J(p)}\]Z\ ,\label{commu1}\ee  where we have omitted  the term $\int {d^4p\o (2\pi)^4}\bar\u_{in,J}(p)C_I^J(p){\d\o\d {\tilde  \zeta}_I(p)}Z$ since ${\d\o\d {\tilde  \zeta}_I(p)}Z$ is a regular function of $p^2$  at $p^2=m^2_g$ for any ghost mass $m_g$ and the momentum integral in Eq.(\ref{commu1}) is restricted to the union of ghost mass-shells where a pole in ${\d\o\d {\tilde  \zeta_I}(p)}Z$ should be needed in order to compensate the zeros of the ghost wave operator $C$.
For the same reason the functional derivative ${\d\o\d {\tilde  \c}_\b(p)} Z$ appearing in the same integral should be replaced with its mass-shell singular part, which, using  Eq.(\ref{d2}), is easily seen to be given by  $\C^\b_I(p) {\d\o\d \tilde  {\bar\s}_I(p)} Z$.
Indeed one finds that the two-point function ${\d^2\o\d\s^I(0)\d\tilde\c_\a(p)}Z_c|_{{\cal J}=0}$ is equal to $\C^\a_J(p) {\d^2\o\d\s^I(0)\d\tilde{\bar\s}_J(p)}Z_c|_{{\cal J}=0}$ where the second factor is the ghost propagator.

Furthermore, from Eq.(\ref{eb}) one has that ${\d\o\d {\tilde  J}^I(p)}Z$ should be replaced with $-{1\o\xi}V^\a_I(-ip){\d\o\d {\tilde  j}^\a(p)}Z$.
Hence, taking into account Eq.(\ref{sti2}), Eq.(\ref{commu1}) should be written:
\bea [{\cal S}, \Sigma_B]Z=-\int {d^4p\o (2\pi)^4}\[\f_{in}^\a(p)\C_{\b\a}(-p)\C^\b_I(p) {\d\o\d  \tilde { \bar\s}_I(p)}\acca-{1\o\xi}\u_{in}^I(p)C_I^J(-p)V_J^\a(-ip){\d\o\d {\tilde  j}^\a(p)}\]Z\nn=
\int {d^4p\o (2\pi)^4}\[\u_{in}^I(p)\C_I^\a(-p)\bar\C_{\b\a}(-p){\d\o\d {\tilde  j}_\b(p)}\acca -{1\o\xi}\f_{in}^\a(p)V_\a^{J} (ip)C_I^J(p) {\d\o\d  \tilde { \bar\s}_I(p)} \]Z\ .\label{commu2}\eea
We have used the relation $\C_{\a\b}(p)= \C_{\b\a}(-p) $. 
This result is the kernel of the first rigorous unitarity proof given in \cite{brs}. Here we shall follow the simpler analysis given in \cite{ko}.

Following Kugo-Ojima, we introduce an operator $Q$ acting on the asymptotic state space and annihilating the vacuum state: $Q|0\rangle=0$.
The new operators satisfies the following relations
\bea   \[Q, \f^\a_{in}(p)\]=-i\ \C_I^\a (-p)\u^I_{in}(p)  \nn
\{Q, \ub_{in, I}(p)\}=i {1\o\xi}V_\a^{J} (ip)\f_{in}^\a(p)=-ib^I_{in}(p)\ .\label{com}\eea
We remind that with our conventions all the asymptotic 
fields are Hermitian  except $\ub_{in, I}$
which is anti-Hermitian, and the gauge fixing action is Hermitian. The operator Q generates a nilpotent transformation on the Fock space. Its kernel $KerQ$ is the
subspace generated by the action of the vacuum state of the positive frequency (creation) part of the physical fields  and by $b_{in,
I}^{(+)}$ and $\u_{in}^{(+)I} $, while its image $ImQ$ is the subspace generated by
$b_{in, I}^{(+)}$ and $\u_{in}^{(+)I } $. 

Now it is interesting to define the cohomology of $Q$, that is, the quotient space $KerQ/ImQ$. This is the linear space of the $Q$-equivalence classes of the elements of $KerQ$. We consider $Q$-equivalent two elements of $KerQ$  if their difference belongs to $ImQ$. It is easy to verify that the cohomology of $Q$ coincides with $Q$-equivalence classes of the states of the subspace of the Fock space generated by  the action on the vacuum state of the positive frequency  part of the physical fields.
Notice  that the asymptotic properties of these field components are completely
determined by the invariant part of $\C$ and hence it is expected that they generate
a positive norm space. It is also a direct consequence of the nilpotency of $Q$ that the
states in the image of $Q$ have vanishing scalar product with those in  its
kernel, they are in particular zero norm states.

Using (\ref{com}) one has \be i \[Q,\Sigma_B\] Z=[{\cal S}, \Sigma_B]Z\ .\ee It follows that:
\bea \[Q,\S\] =\[Q,:e^{\Sigma_B}:\]  Z|_{{\cal J}=0}= :\[Q, e^{\Sigma_B} \]:  Z|_{{\cal J}=0}=-i [{\cal S}, :e^{\Sigma_B}:] Z|_{{\cal J}=0}\nn=i:e^{\Sigma_B}:{\cal S}Z|_{{\cal J}=0}=0\ .\label{qs}\eea
Notice that  $Q$ acts separately on the positive   and the negative frequency parts of the asymptotic fields and this justifies the second identity in Eq.(\ref{qs})
which shows that the commutator $\[Q,\S\]$ vanishes.
%\bea :\[Q,e^\Si\]=:\[Q,e^\Sigma\]:Z|_{{\cal J}=0}
%=\nn -i:e^\Sigma\(b_{in, J} 
%C_{(asy)I}^J \p_{{\bar \sigma}_I}+
%\u^I_{in} C_{(asy)I}^K\p_{J^K}\):Z|_{{\cal J}=0}=0\
%.\label{61}\eea
Under the assumption that the measure corresponding to a Hermitian Lagrangian
defines a "pseudo"uni- tary $\S$-matrix in the asymptotic Fock space:
\be \S \S^{\dag} =\S^{\dag} \S =I\ ,\ee
 and that the physical space is  a positive metric  space - both assumptions are
true in perturbation theory - we conclude that, owing to (\ref{qs}) and to  the
above discussed properties of $Q$, the $\S$-matrix is unitary in the physical space
identified with the cohomology of $Q$. That is: if the initial state $|i\rangle$ is a physical state and hence  it belongs to the kernel of $Q$ and has positive norm, $\S|i\rangle$ belongs to the kernel of $Q$ and has the same norm as $|i\rangle$, therefore it defines a $Q$-equivalence class corresponding to a positive norm physical state.

We consider for example an $SU(2)$ Higgs model in the tree approxima-
tion \cite{brs}. This model involves an iso-triplet of vector
fields ${\vec A}_\mu$, a triplet of Goldstone particles ${\vec \pi}$, and the Higgs
field $\sigma$ that appears in our calculations only through its vacuum expectation value $ V$. Therefore now the symbol $\f^\a$ corresponds to 
${\vec A}_\mu$ and ${\vec \pi}$. We add a further iso-triplet of Nakanishi-Lautrup
multipliers $\vec b$.
In the tree approximation the free Lagrangian density is given by:
\be{\cal L}=-{\vec F_{\mu\nu}\cdot\vec F^{\mu\nu}\o 4} + {1\o2}\(\p\vec \pi-gV\vec
A\)^2+\vec b\cdot\(\p\vec A+\rho\vec\pi\)+\xi{b^2\o 2}-\vec{\bar\u}\cdot(\p^2+g\rho V)\vec\u\ ,\ee
From now on we shall disregard the isotopic indices since all the wave operators are
diagonal in the isotopic space.
The wave matrix  $\C$ defined in (\ref{d1}) is given by:
 \be\label{hig2}
\left(\ \ \ber{cc}\ \ p^\mu p^\nu-g^{\mu\nu}p^2+g^2V^2g^{\mu\nu} & -igVp^\mu  \\
igVp^\nu &
 p^2 
 \eer\ \right),\ee
where the first row and line correspond to $A$ and the second ones to $\pi$.
The matrix $V^\a_I$ defining the gauge fixing in (\ref{lingau}) corresponds to:
\be\label{hig3}
\left(\ \ \ber{c}\ -ip^\mu \\ \rho
\eer\ \right),\ee
Notice that the isotopic indices are hidden. The ghost wave operator $C(p)$ is given by $ p^2-g\rho V$.

The wave operator of the gauge and scalar fields $\bar\C_{\a\b}(-p)$ appearing in Eq. (\ref{siste2}) is:
\be\label{hig4}
\left(\ \ \ber{cc}\ \ p^\mu p^\nu(1-{1\o\xi})-g^{\mu\nu}p^2+g^2V^2g^{\mu\nu} & i(gV-{\rho\o\xi})p^\mu  \\
-i(gV-{\rho\o\xi})p^\nu &
 p^2 -{\rho^2\o\xi}
 \eer\ \right)\ .
 \ee
Forgetting the isospin degeneracy its determinant is given by:
\be\det|\bar\C|=-{1\o\xi}(p^2-(gV)^2)^3(p^2-g\rho V)^2\ ,\label{hig5}\ee
however Eq.(\ref{siste2}) has only four independent solutions, three of them correspond  to $p^2=m_p^2=(gV)^2$, the remaining one  corresponding to
$p^2=m_g^2=g\rho V$ where   the above determinant has a double zero. Looking at the solutions of Eq.(\ref{siste2}) and considering in particular the {\it longitudinal} ones in which $A^{(l)}_\mu(p)\sim p_\mu$, one sees that  for a generic choice of $\rho$ there is a single solution which, in contrast with the hypothesis made at  the beginning of this chapter, corresponds to a dipole singularity  of  the gauge and scalar field propagator  which is  degenerate with the simple pole of the ghost propagator. This is by no means surprising since the asymptotic longitudinal vector field satisfies the free field equations $(p^2-\xi gV)\vec {A^{(l)}}_\mu(p)=ip_\mu(\xi gV-\rho)\vec\pi$ which is analogous to the field equation in   Landau's gauge QED. One can find some details on the structure of the asymptotic state space  in Appendix C and in  Landau's gauge QED in \cite{scholar}. The dipole singularity disappears and one finds a fifth independent solution of  Eq. (\ref{siste2}) mass degenerate with the ghosts if $\rho=\xi gV$. This is called the {\it special} 't Hooft choice. It is easy to verify that  Eq. (\ref{siste2}) has solutions with $b\not\equiv 0$ only in this special case.

The matrix $\C^I_\a$ defined in (\ref{d2}) in the tree approximation is:
\be\label{hig6}
\left(\ \ \ber{c}\ ip^\mu \\ gV
\eer\ \right)\ .\ee It is apparent that its columns correspond to solutions of  Eq. (\ref{siste2}) on the ghost mass-shell in agreement with Eq. (\ref{g2}).
 The $Q$ operator is defined by the conditions:
 \bea   \[Q,\vec A^\mu_{in}(p)\]=-\ p^\mu \vec\u_{in}(p)  \quad \ ,\quad  \[Q,\vec \pi_{in}(p)\]=-i gV  \vec\u_{in}(p) \nn
\{Q, \vec{\bar\u}_{in}(p)\}={1\o\xi}[p_\mu A^\mu_{in}(p) +i\rho \vec\pi_{in}(p)]\equiv -ib_{in}(p)\  .\label{com2}\eea
This operator is apparently nilpotent since on the ghost mass-shell \be\{Q, \[Q,\vec A^\mu_{in}(p)\]\}=0\ .\ee

In order to have a look at the physical content of the theory we introduce three space-like polarization vectors $\epsilon_a^\mu(p)$ orthogonal to the 
momentum $p$ and such that $\epsilon_a^\mu(p)\epsilon_{b\nu}=-\d_{ab}$. Due to Eq.(\ref{siste2}) the asymptotic field $\vec A^\mu_{in}(p)=\sum_a \vec \f_{a, in}(p)\epsilon_a^\mu(p)$ has support on the mass-shell $p^2=g^2V^2$ and its positive frequency part generates a positive norm subspace of the asymptotic space.  Furthermore  from the above relation it is apparent that $[Q,\f_{a, in}(p)]=0$ and hence the $Q$-equivalence classes of these positive norm states identify the cohomology of $Q$, that is the physical state space.

Notice that the presence of dipole singularities remarked above, which, as shown in Appendix C, corresponds to ghost-degenerate unphysical states with opposite norm, does not affect the conclusions of our analysis \cite{brs}\cite{scholar}.

 \sp
\appendix
\sec{Appendix}
The differential system $\{X\}$ being integrable one can choose local charts of
coordinates trivializing the fibration; let us indicate by $\{\xi\}$ the gauge
invariant coordinates, that are constant  along the orbits, and by $\{\eta\}$ those
parametrizing points on the orbits; the transition functions between two neighbouring
charts are:
\be \xi_a=\xi_a\(\xi_b\)\quad\ ,\quad\n_a=\n_a\(\xi_b,\n_b\)\ .\label{changevar}\ee
In a given chart the elements of the differential system are:
\be X_I\equiv X_I^J\( \xi,\n\)\p_{\n^J}\ .\ee
On every fibre we define the adjoint system of differential 1-forms according
\be \u^I\equiv\u^I_J\( \xi,\n\)d\n^J\ ,\ee
with:
\be X_I^L\u_L^J=\d_I^J\ .\ee
Taking into account the commutation relation (\ref{9}) one verifies directly that
these 1-forms satisfy the Maurer-Cartan equation:
 \be d_V\ \u^I\equiv d\n^L\p_{\n^L}\
\u^I=-{1\o2} C^I_{JK}\ \u^J\u^K\ .\ee
The above equations  define explicitly and, according to (\ref{changevar}), 
globally on very orbit, the exterior
algebra involved into the definition of the Faddeev Popov measure. However they cannot
be directly translated into their field theory equivalent due to locality. Indeed, even
if the differential system  $\{X\}$ is given as a set of local functional differential
operators in the gauge field variables, the trivializing coordinates are non-local
with respect to the fields.

 This difficulty is overcome replacing the system of
generators of the vertical exterior algebra $\{d\n\}$ with $\{\u\}$ that in this way
appear into the theory as new Grassmannian local field variables. This however
requires that the vertical exterior derivarive $d_V$ be written according (\ref{15}):
\be d_V=\u^I X_I-{1\o2} C^I_{JK} \ \u^J\u^K \p_{\u^I}\ ,\ee
where the first term in the right-hand side accounts for the action of $d_V$ on
functions while the second term acts on the exterior algebra generators.

 \sp

\sec{Appendix}
In this appendix we prove (\ref{1a}) exploiting (\ref{21}).
The first step will be the proof of two lemmas whose recursive use will lead to
(\ref{1a}).
Let us  consider the set of cells $\{U_a\}$ and the corresponding
partition of unity $ \{\chi_a\}$ and gauge fixing functionals $\{\T_a\}$
We define:
\be \(s\chi_{a_{1}}...s\chi_{a_{n-1}}\ \chi_{a_{n}}\)_A\equiv
\sum_{k=1}^n(-1)^{k-n}\chi_{a_{k}}s\chi_{a_{1}}...s\check\chi_{a_k}...s\chi_{a_{n}}\
,\label{a1}\ee
where the check mark above $\chi_{a_k}$ means that the corresponding term should be
omitted. It is fairly evident that the functional (\ref{a1}) is antisymmetric with
respect to permutations of the indices $\( a_1,..., a_n\)$ and that its support  is
contained in the intersection of the corresponding cells.

It is apparent that:
\be s \(s\chi_{a_{1}}...s\chi_{a_{n-1}}\
\chi_{a_{n}}\)_A=(-1)^{n+1}n\ s\chi_{a_{1}}...s\chi_{a_{n}}\ .\label{a2}\ee
Furthermore, taking into account that $ \{\chi_a\}$ is a partition of unity on the support of the functional measure, one
has: \bea\sum_{a_{n+1}}\(s\chi_{a_{1}}...s\chi_{a_{n}}\ \chi_{a_{n+1}}\)_A
=\sum_{a_{n+1}}\sum_{k=1}^{n+1}(-1)^{k-n-1}\chi_{a_{k}}
s\chi_{a_{1}}...s\check\chi_{a_k}...s\chi_{a_{n+1}}\nn
=s\chi_{a_{1}}...s\chi_{a_{n}}\ ,\label{a3}\eea
indeed only the term with $k=n+1$ contributes to the second member giving the
right-hand side of this equation. Comparing (\ref{a2}) with  (\ref{a3}), we
get:
\be s \(s\chi_{a_{1}}...s\chi_{a_{n-1}}\
\chi_{a_{n}}\)_A=(-1)^{n+1}n\ \sum_{a_{n+1}}
\(s\chi_{a_{1}}...s\chi_{a_{n}}\ \chi_{a_{n+1}}\)_A\ .\label{a4}\ee
Let now $A_{a_{1},...,a_{n+1}}$ be antisymmetric in its indices, one has:
\bea \sum_{a_{1},...,a_{n+1}}A_{a_{1},...,a_{n+1}}\p_{\T}\(\T_{a_{1}}...\T_{a_{n}}\)
e^{is\T_{a_{1},...,a_{n}}}\nn =i{(-1)^n\o
n+1}\sum_{a_{1},...,a_{n+1}}A_{a_{1},...,a_{n+1}}\ s\
\p_{\T}\(\T_{a_{1}}...\T_{a_{n+1}}\) e^{is\T_{a_{1},...,a_{n+1}}} \ .\label{a5}\eea
Indeed, using the identity:
\be\sum_{a_{1},..,a_{n+1}}A_{a_{1},..,a_{n+1}}V_{a_{1},..,a_{n}}={(-1)^n\o
n+1}\sum_{k=1}^{n+1}(-1)^{k+1}\sum_{a_{1},..,a_{n+1}}A_{a_{1},..,a_{n+1}}V_{a_{1},..,
\check a_k,..,a_{n+1}}\ ,\ee
 the left-hand side of (\ref{a5}) is written:
\bea{(-1)^n\o
n+1} 
\sum_{a_{1},...,a_{n+1}}A_{a_{1},..,a_{n+1}}\sum_{k<l=1}^{n+1}
(-1)^{k+l}\T_{a_{1}}..\check \T_{a_{k}}..\check \T_{a_{l}}..\T_{a_{n+1}}\nn
\(e^{is\T_{a_{1},..,\check a_k,..,a_{n+1}}}-e^{is\T_{a_{1},..,\check a_l,..,a_{n+1}}}\)
 \nn =i{(-1)^n\o
n+1} 
\sum_{a_{1},...,a_{n+1}}A_{a_{1},..,a_{n+1}}\sum_{k<l=1}^{n+1}
(-1)^{k+l}\T_{a_{1}}..\check \T_{a_{k}}..\check \T_{a_{l}}..\T_{a_{n+1}}\nn
s\(\T_{a_k}-\T_{a_l}\)e^{is\T_{a_{1},...,a_{n+1}}}\ ,
 \eea
from which one reaches (\ref{a4}).

Let us now consider the extension of (\ref{sti}) to the case of a cell decomposition of
$\fc$, one has:
\bea \int d\mu_C\sum_{a}\chi_{a}e^{is\T_{a}}sX=-\int
d\mu_C\sum_{a_1}s\chi_{a_1}e^{is\T_{a_1}}X\nn=\int d\mu_C\sum_{a_1,a_2}
\(s\chi_{a_{1}}\chi_{a_{2}}\)_Ae^{is\T_{a_1}}X\nn={1\o2}\int d\mu_C\sum_{a_1,a_2}
\(s\chi_{a_{1}}\chi_{a_{2}}\)_A\(e^{is\T_{a_1}}-e^{is\T_{a_2}}\)X\nn=
-{i\o2}\int d\mu_C\sum_{a_1,a_2}
\(s\chi_{a_{1}}\chi_{a_{2}}\)_As\p_{\T}\(\T_{a_{1}}\T_{a_{2}}\)e^{is\T_{a_1,a_2}}X
\ .\label{a6}\eea
In (\ref{a6}) we have used (\ref{a3})  and (\ref{a5}) with $n=1$. From (\ref{a6}) we
have: \bea \int d\mu_C\[\sum_{a}\chi_{a}e^{is\T_{a}}+{i\o2}\int d\mu_C\sum_{a_1,a_2}
\(s\chi_{a_{1}}\chi_{a_{2}}\)_A\p_{\T}\(\T_{a_{1}}\T_{a_{2}}\)e^{is\T_{a_1,a_2}}
\]sX\nn=-{i\o2}\int d\mu_C\sum_{a_1,a_2}
s\(s\chi_{a_{1}}\chi_{a_{2}}\)_A\p_{\T}\(\T_{a_{1}}\T_{a_{2}}\)e^{is\T_{a_1,a_2}}\
.\eea
 Using recursively the same equations we get: 
\bea   d\mu_C\[\sum_a\chi_ae^{is\T_{a}}-
\sum_{n=1}^{m}i^n{(-1)^{n(n-1)\o2}\o n+1}\(s\chi_{a_{1}}...s\chi_{a_{n}}\
\chi_{a_{n+1}}\)_A
\r.\nn\l.\p_{\T}\(\T_{a_{1}}...\T_{a_{n+1}}\)e^{is\T_{a_{1},...,a_{n+1}}}\]sX  \nn=
i^m\ {(-1)^{m(m-1)\o2}\o m+1}\int d\mu_Cs\(s\chi_{a_{1}}...s\chi_{a_{m}}\
\chi_{a_{m+1}}\)_A\nn\p_{\T}\(\T_{a_{1}}...\T_{a_{m+1}}\)s\
e^{is\T_{a_{1},...,a_{m+1}}}X \ .\label{a7}\eea
It is clear that, if the maximum effective number of intersecting cells is $N$ the
right-hand side of (\ref{a7}) vanishes for $m\geq N$.

\sec{Appendix}

 The aim of this Appendix is to clarify the structure of the Fock space associated with a generalized free  Hermitian field $\f$ whose propagator presents a dipole singularity, that  is such that:
\be \langle0|T(\f(x)\f(0))|0\rangle=i\int {dp\o(2\pi)^4}{e^{ipx}\o(m^2-p^2-i0+)^2}\ .\ee
Using the Lehmann spectral representation it follows that the Wightman function is
\be \langle0| \f(x)\f(0) |0\rangle=\int {dp\o(2\pi)^4}e^{ipx}\theta(p^0)\d'( p^2-m^2) \ ,\label{lehm}\ee and hence, if the field is Hermitian, the corresponding Fock space must be an indefinite metric space. Indeed $\d'(x)$ is not a positive distribution.  We shall call pseudo-Hermitian an Hermitian operator in an indefinite metric space and we shall label the pseudo-Hermitian conjugate by a dagger. 
It is clear that our field satisfies the linear equation:
\be (\p^2+m^2)^2\f(x)=0\ ,\ee whose general solution is
\be\f(x)=\int {d p\o(2\pi)^2}\ e^{ipx}[\a(p)\d( p^2-m^2) +\b(p)\d'( p^2-m^2) ]\ .\label{solu}\ee
Here $\a(p)$ and $\b(p)$ have analogous $p\equiv (p^0\ ,\v p)$ dependence, in particular, $\a(p)=\a_+(\v p)\theta(p^0)+\a_-(\v p)\theta(-p^0)$. The stability condition for the vacuum state with our metric choice, $px=p^0x^0-\v p\cdot\v x$, implies that $\a_-(\v p)|0\rangle=0$ and $\b_-(\v p) |0\rangle=0$. Thus $\a_-$ and $\b_-$ are annihilation operators. An alternative expression for the field is given integrating over $p^0$ and taking into account that, if the function $f(x)$ continuous with its derivative has $n$ non-degenerate zeros $x_i$ one has:
\be\d(f(x))=\sum_{i=1}^n{\d(x-x_i)\o|f'(x_i)|}\quad {\rm and} \quad \d'(f(x))={1\o f'(x)}\sum_{i=1}^n {\d'(x-x_i)\o|f'(x_i)|}  \ .\ee Setting $E(\v p)=\sqrt{|\v p|^2+m^2}$ one gets:
\bea\f(x)=\int {d p\o(2\pi)^22E(\v p)}\ e^{ipx}\[\a_+(\v p)\d( p^0-E(\v p))+\a_-(\v p)\d(p^0+E(\v p)) \acca+{\b_+(\v p)\o2p^0}\d'( p^0-E(\v p)) +{\b_-(\v p)\o2p^0}\d'( p^0+E(\v p)) \]\nn=\int {d\v p\o(2\pi)^22E(\v p)}e^{ipx}|_{p^0=E(\v p)}
[\a_+(\v p)+{\b_+(\v p)\o 2E^2(\v p)}(1-ix^0 E(\v p)) ]\nn+\int {d\v p\o(2\pi)^22E(\v p)}e^{ipx}|_{p^0=-E(\v p)}
[\a_-(\v p)+{\b_-(\v p)\o 2E^2(\v p)}(1+ix^0 E(\v p)) ]\ .
\eea Therefore we see that the field is pseudo-Hermitian if
\be \a_+^\dag(\v p)=\a_-(-\v p)\quad\  ,\quad\b_+^\dag(\v p)=\b_-(-\v p)\ .\ee

Writing Eq(\ref{lehm}) in terms of $\a(p)$ and $\b(p)$ one gets from Eq.(\ref{solu})
\bea  \langle0| \f(x)\f(y) |0\rangle = \int {d^4p\o(2\pi)^42E(\v p)}e^{ip(x-y)}{\d'( p^0-E(\v p)) \o2p^0}\\&&=
\int {d\v p\o(2\pi)^42E^3(\v p)}e^{ipx}|_{p^0=E(\v p)}
(1-i(x^0-y^0) E(\v p)) \nn=\int {d\v p\ d\v q\o(2\pi)^44E(\v p)E(\v q)}e^{i(px+qy)}|_{p^0=E(\v p), q^0=-E(\v q)}\nn\langle0|\(\a_+(\v p)+{\b_+(\v p)\o 2E^2(\v p)}(1-ix^0 E(\v p))\)\(\a_-(\v q)+{\b_-(\v q)\o 2E^2(\v q)}(1+iy^0 E(\v q))\)|0\rangle\nonumber\ ,
\eea from which one finds the commutation conditions:
\bea [\a_+(\v p),\b_-(\v q)]=[\b_+(\v p),\a_-(\v q)]=2E(\v p)\d(\v p+\v q)\nn  [\a_+(\v p),\a_-(\v q)]=-E^{-1}(\v p)\d(\v p+\v q)\quad  [\b_+(\v p),\b_-(\v q)]=0\ .\label{comrel}\eea Taking into account the pseudo-Hermiticity conditions, these commutation  relations can be {\it diagonalized} introducing the annihilation and creation operators
$A_\s(\v p)$ and $A^\dag_\s(\v p)$ with $\s=\pm$ defined by:
\be A_\s(\v p)\equiv \sqrt{2\o\sqrt{5}}( E(\v p)\a_-(\v p)+\s){\sqrt{5}+\s 1\o 4E(\v p)}\b_-(\v p)\ ,\ee whose commutation rules are:\be[A_\s(\v p), A^\dag_{\s'}(\v q)]=\d_{\s,\s'}\s2 E(\v p)\d(\v p-\v q)\quad
[A_\s(\v p), A_{\s'}(\v q)]=o\ . \ee
It is apparent that the corresponding Fock space has indefinite metric corresponding to the operator
$(-1)^{N_-}$ with $N_-=\int d\v p/(2E(\v p))A^\dag_-(\v p)A _-(\v p)\ .$

%\section{Bibliography}

\end{document}